\documentclass{PoS}
\title{Charm quark mass dependence in the CTEQ NNLO global QCD analysis}

\ShortTitle{Charm quark mass dependence in the CTEQ NNLO global QCD analysis}

\author{Jun Gao,$^a$ \speaker{Marco Guzzi}$^{b}$
and Pavel M. Nadolsky$^{a}$\\
\llap{$^a$}Department of Physics, Southern Methodist, University\\
Dallas, TX 75275, USA \\
\llap{$^b$}Deutsches Elektronen-Synchrotron DESY\\
Notkestrasse 85, 22607 Hamburg, Germany\\
E-mail: \email{jung@mail.smu.edu}, \email{marco.guzzi@desy.de},
\email{nadolsky@physics.smu.edu}}

\abstract{
We discuss the impact of the charm quark mass in the CTEQ NNLO global 
analysis of parton distribution functions of the proton. 
The $\overline{\rm MS}$ mass $m_c(m_c)$ of the charm quark 
is extracted in the S-ACOT-$\chi$ heavy-quark 
factorization scheme at ${\cal O}(\alpha_s^2)$ accuracy 
and found to be in agreement with the world-average value.
Impact on $m_c(m_c)$ of combined HERA-1 data on 
semiinclusive charm production at HERA collider 
and contributing systematic uncertainties are reviewed.}
 
\FullConference{XXI International Workshop on Deep-Inelastic Scattering and Related Subjects\\
                 22-26 April, 2013\\
                 Marseilles, France}

\begin{document}

{\bf Introduction.}
Measurements of lepton-nucleon deep-inelastic scattering (DIS) cross
sections become increasingly sensitive to scattering of heavy quarks,
$c$ and $b$, at energies comparable to heavy-quark masses. This
progress motivated several recent analyses 
\cite{Martin:2010db,Abramowicz:1900rp,Alekhin:2012vu}
to determine the mass $m_c$ of the charm quark from the DIS and other
hadronic data in fits of PDFs in the nucleon. 
The functional form of the PDFs preferred by the QCD data is 
dependent on the method by which heavy-quark masses are included in
DIS structure functions \cite{Tung:2006tb}. Consequently 
various precision measurements at the LHC are dependent 
on the heavy-quark treatment in DIS experiments. 

Combined cross sections on inclusive DIS and semiinclusive DIS 
charm production at the $ep$ collider HERA
\cite{Aaron:2009aa,Abramowicz:1900rp} have the best potential 
to constrain the charm mass. On the theory side, 
perturbative QCD (PQCD) calculations for neutral-current DIS 
exist at the 2-loop level in $\alpha_{s}$ both for massless
\cite{SanchezGuillen:1990iq,vanNeerven:1991nn,Zijlstra:1991qc} and massive 
\cite{Laenen:1992zk,Riemersma:1994hv,Harris:1995tu} quarks, 
while massless \cite{Moch:2004xu,Vermaseren:2005qc}
and some massive \cite{Blumlein:2006mh}
coefficient functions were also obtained at the 3-loop level. 
With such accuracy, it is possible to determine the charm quark
mass and its uncertainty from the DIS data. 
In Ref.~\cite{Gao:2013wwa} we explored constraints on the 
$\overline{\textrm{MS}}$ charm mass $m_c(m_c)$ 
in the CT10 NNLO PDF analysis in order to compare them to  
$m_c$ determinations from non-DIS experiments and by other groups. 
This study examined the feasibility of the $m_c$
extraction from DIS measurements, which are unique in their right
as spacelike charm production processes. The other goal was to
determine $m_c(m_c)$ in a General Mass Variable Flavor Number (GM-VFN) scheme 
S-ACOT-$\chi$ \cite{SACOTchi}, the default heavy-quark scheme of CT
analyses. This scheme is well-suited for theoretical expolaration of
factors affecting the determination of $m_c(m_c)$, as a result of its close
connection to the QCD factorization theorem \cite{Collins:1998rz} 
for DIS with heavy quarks. Recently, the S-ACOT-$\chi$ calculations
wer extended to ${\cal O}(\alpha_s^2)$, or NNLO, in NC DIS
\cite{Guzzi:2011ew}, which significantly reduced theoretical
uncertainties compared to the previously employed
\cite{Abramowicz:1900rp} NLO S-ACOT calculations.  

{\bf Implementation of the $\overline{\textrm{MS}}$ mass.} 
Our calculation ~\cite{Gao:2013wwa} takes $\overline{\rm MS}$ quark masses
as the input for the whole calculation. The transition from the 3-flavor
to 4-flavor evolution in $\alpha_{s}$ and PDFs occurs
at the scale equal to this input mass.
The massive 2-loop coefficient functions for neutral-current DIS with
explicit creation of $c\bar{c}$ pairs \cite{Riemersma:1994hv} and the operator
matrix elements $A_{ab}^{(k)}$ \cite{Buza:1996wv} that we use
require the pole mass as their input. For these parts, the $\overline{\rm MS}$
mass is converted to the pole mass according to the 2-loop perturbative
relation in Eq. (17) of \cite{Chetyrkin:2000yt}.
The global fit is sensitive to the number of loops included in
$\overline{\rm MS}$ conversion. We explore this sensitivity 
by implementing two methods. In the first method, the $\overline{\rm
  MS}$ mass is converted to the pole mass by the 2-loop 
relation in both ${\cal O}(\alpha_s)$ and ${\cal O}(\alpha_s^2)$ 
radiative contributions to heavy-quark coefficient functions. 
In the second method, the 2-loop (1-loop) conversion is performed in 
the ${\cal O}(\alpha_s)$ and ${\cal O}(\alpha_s^2)$ terms in the 
Wilson coefficient functions and OME's, respectively. This is argued
to be equivalent to calculating DIS structure functions directly  
in terms of the $\overline{{\rm MS}}$ mass and improve
perturbative convergence of the best-fit values 
for $m_{c}(m_c)$~\cite{Alekhin:2010sv}.

{\bf Theoretical inputs.} 
Several aspects of the QCD calculation affect determination of $m_c$.   
In a comprehensive factorization scheme such as GM-VFN, the exact 
charm mass enters hard matrix elements for charm particle creation (FC) 
in the final state, such as 
$\gamma^* g \rightarrow \bar{c}c$ in NC DIS. At the same time, GM-VFN 
introduces several energy scales that are {\it approximately} equal to the
charm mass, including the switching scale between 3 and 4-active
flavors and the effective mass in the flavor-excitation (FE) matrix elements 
(with incoming heavy quarks). Our analysis indicates that it is the
exact $m_c(m_c)$ in the FC cross sections, and not the approximate
mass scales, that primarily controls the agreement with the DIS
data. 

GM-VFN schemes used in the PDF fits \cite{SACOTchi,Thorne:1997ga,Thorne:2006qt,Forte:2010ta}
differ primarily in the form of approximation for FE 
coefficient functions at $Q$ comparable to $m_{c}$, due to powerlike
contributions $(m_{c}^{2}/Q^{2})^{p}$ with $p>0$ arising near the threshold. 
In S-ACOT-$\chi$ the form of these contributions is selected based on
the general consideration of energy-momentum 
conservation (reference 3 in \cite{SACOTchi}). Their detailed form can
be varied to estimate the associated higher-order uncertainty 
in the extracted $m_{c}$ by introducing
a generalized rescaling variable $\zeta$ ~\cite{Nadolsky:2009ge}, 
implicitely defined by
$x=\zeta\, \left(1+\zeta^{\lambda}M_{f}^{2}/Q^{2}\right)^{-1}$. The
default (and best motivated) form of the rescaling variable is
obtained assuming $\lambda=0$. However, other values of $\lambda$
between 0 and 1 can be used to estimate the uncertainty.

\begin{table}[ht]
\begin{centering}
\begin{tabular}{c|c|c|c|c}
\hline 
Theoretical systematic uncertainty   & DIS scale  & $\alpha_{s}(M_{Z})$  & $\lambda$  & $\chi^{2}$ definition \tabularnewline
\hline 
Parameter range  & $[Q/2,\ 2Q]$  & {[}0.116,\, 0.120{]}  & {[}0, 0.2{]}  & -- \tabularnewline
\hline 
$\delta m_c(m_c)$ (GeV)  & ${}^{+0.02}_{-0.02}$  & ${}^{+0.01}_{-0.01}$  & ${}^{+0.14}_{-0}$  & ${}^{+0.06}_{-0}$ \tabularnewline
\hline 
\end{tabular}
\par\end{centering}
\caption{\label{unc} Shifts of the optimal value of the charm mass $m_{c}(m_{c})$
obtained by varying theoretical inputs.}
\end{table}

Theoretical uncertainties are summarized in Table \ref{unc}, showing 
shifts in the extracted $m_c(m_c)$ due to the factorization/renormalization
scale in DIS cross sections, 
$\alpha_{s}(M_{Z})$, the $\lambda$ parameter in the rescaling
variable, and implementation of experimental correlated systematic
errors. The last source of uncertainty arises from the
existence of several prescriptions (designated as ``extended T'' and ``D'' methods in Ref.~\cite{Gao:2013xoa}) 
for including correlated systematic errors from the fitted experiments 
into the figure-of-merit function $\chi^2$.

\begin{figure}[ht]
\includegraphics[width=0.53\columnwidth]{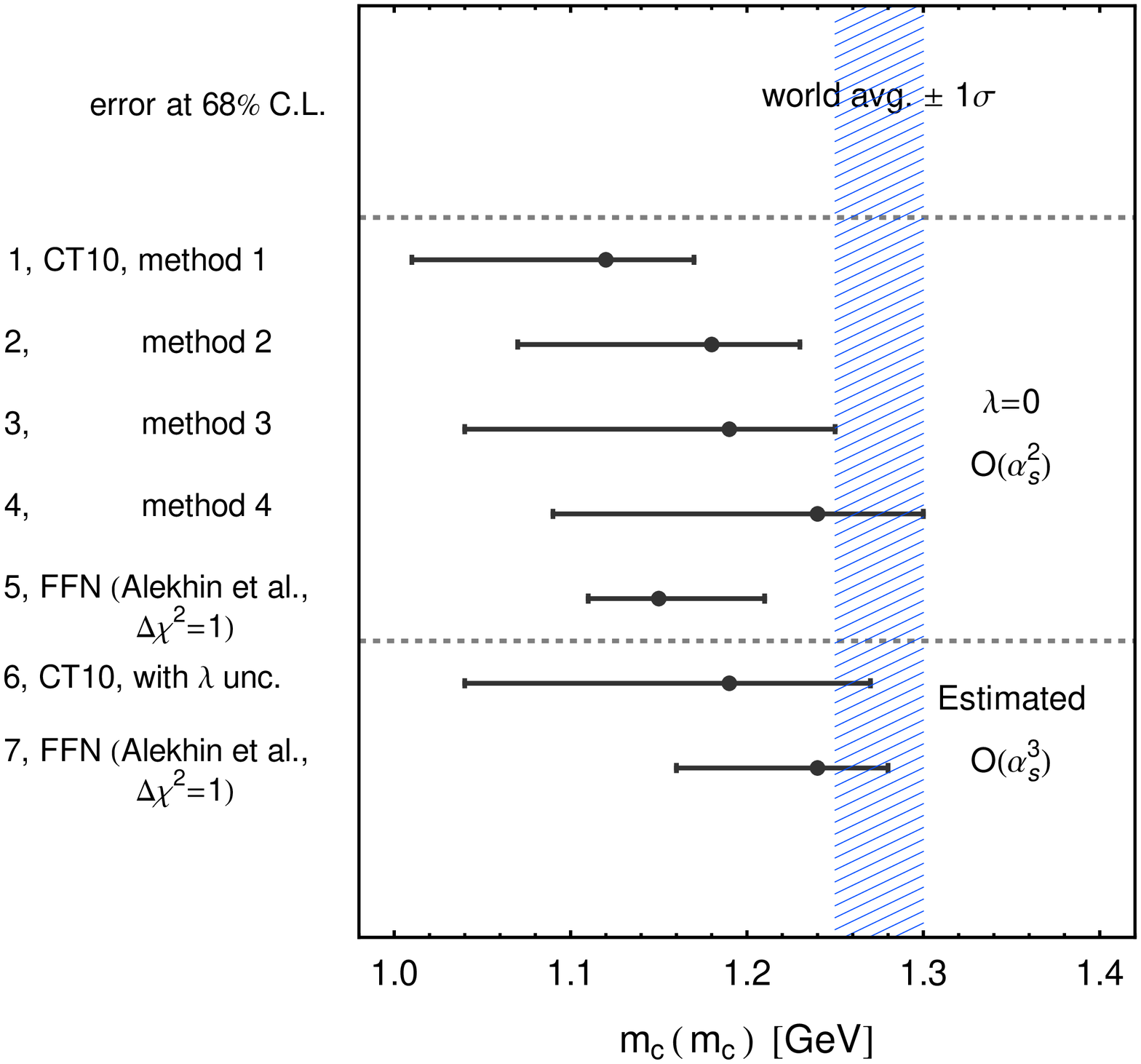}
\includegraphics[width=0.47\columnwidth]{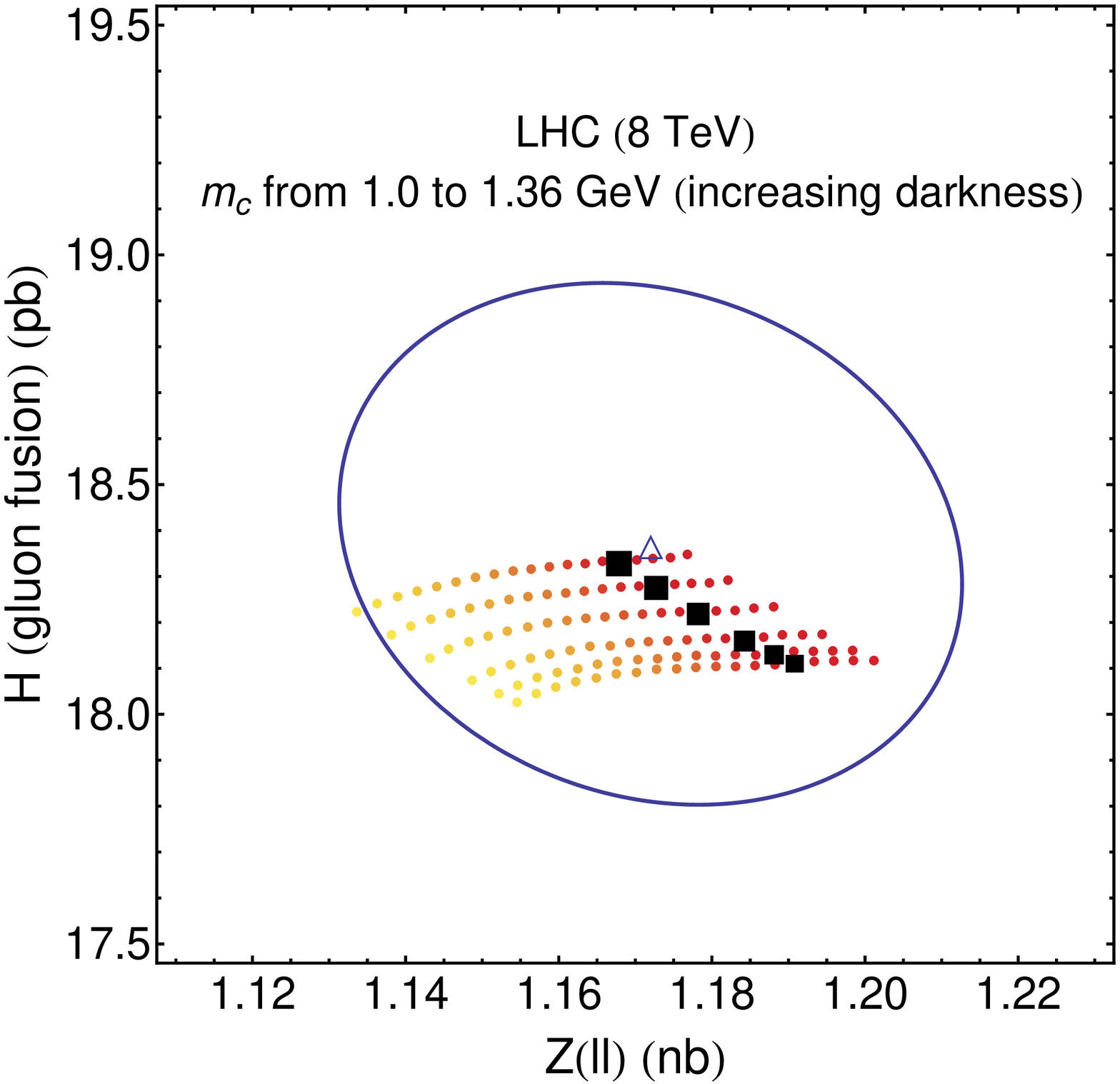} 
\caption{\label{fig:bench} (a) Best-fit values of $m_{c}(m_{c})$ with uncertainties. (b) NNLO cross sections for SM $H^0$ boson
  and $Z^{0}$ boson production at the LHC 8 TeV.}
\end{figure}

{\bf Results of the fit.} 
Our main results are illustrated in Fig.~\ref{fig:bench}, with details
provided in Ref.\cite{Gao:2013wwa}. The left subfigure shows the best-fit
$m_c$ and its uncertainties. At order $\alpha_s^2$, the highest fully implemented order in
our calculation, these values are found with four methods.
Methods 1 and 2 correspond to
the ``extended $T$'' and ``experimental'' $\chi^2$ definitions
respectively~\cite{Gao:2013xoa},   
both using the full $\overline{{\rm MS}}\rightarrow\mbox{pole}$ mass conversion formula, 
and $\lambda=0$. The best-fit values indicated by methods 3 and 4
correspond to the truncated  
mass conversion for the two $\chi^2$ definitions previously mentioned.
The resulting $m_c(m_c)$ values in the four methods are 
$1.12^{+0.05}_{-0.11}$, $1.18^{+0.05}_{-0.11}$, $1.19^{+0.06}_{-0.15}$ and $1.24^{+0.06}_{-0.15}$ GeV,
respectively. Here we quote the  68\% C.L. PDF uncertainties defined
as in the CT10 analysis \cite{Lai:2010vv} based on the value 
of the total $\chi^2$ and agreement with individual experiments. 

As we see, there is some spread in the $m_c$ values depending on the
adopted $\overline{MS}\rightarrow\mbox{pole}$ conversion and $\chi^2$
definition. In addition, moderate dependence exists on the rescaling
parameter $\lambda$, asssociated with missing higher-order corrections.
We can estimate the projected range for the ${\cal O}(\alpha_s^3)$
value of $m_c(m_c)$ by taking the central value found from 
method 3 and adding in quadrature the theoretical
uncertainties obtained by including $\lambda$ dependence. This produces 
$1.19^{+0.08}_{-0.15}$ GeV for the estimated 
${\cal O}(\alpha_s^3)$ value (as shown in line 6 of the left subfigure), 
where the error is computed from the 68\% c.l. contour for $\chi^2$ vs. $\lambda$ 
and adding scale and $\alpha_s$ uncertainties in quadrature.
 
The central $m_c$ is consistent with the PDG value of $1.275\pm 0.025$
GeV within the errors. A tendency of the fits to undershoot the PDG
value may be attributable to the missing ${\cal O}(\alpha_s^3)$
contribution \cite{Alekhin:2012vu}. The results of our fit are
compatible with $m_c(m_c)$  determined from a fit in the fixed-flavor
number (FFN) scheme  \cite{Alekhin:2012vu}, cf. lines 5 and 7 in the
left Fig.~\ref{fig:bench}. However, our PDF error of about 0.15 GeV 
is about twice as large as that quoted in the FFN study. The
reason is that in the FFN analysis the 68\% c.l. PDF 
uncertainty is defined to correspond to $\Delta\chi^{2}=1$ in the total
$\chi^2$, and hence is smaller than the uncertainty according to the
CT10 tolerance criterion. In our analysis, we observe
that the $\chi^2$ dependence on $m_c(m_c)$ is not compatible with the
ideal quadratic dependence required to justify 
the $\Delta\chi^{2}=1$ definition for the $1\sigma$ error. The
actual $\chi^2$ dependence is wider than the quadratic one
and asymmetric, hence the $1\sigma$ error needs to be increased by a factor
of 2-3 compared to its $\Delta\chi^{2}=1$ definition to describe the
observed probability distribution. Besides this difference in the PDF
uncertainty, the results for $m_c(m_c)$ from the S-ACOT-$\chi$ and FFN
fits are in agreement.

Variations in $m_c(m_c)$ impact electroweak cross sections
at the Large Hadron Collider. A plot of NNLO cross sections for Higgs
and $Z^0$ bosons production is shown at 8 TeV for 
$m_{c}(m_{c})$ ranging from 1 to 1.36 GeV and $\lambda=\{0,\ 0.02,\ 0.05,\ 0.1,\ 0.15,\ 0.2\}$.
Darker color corresponds to larger mass values for a fixed $\lambda$.
To access only the uncertainty due to the form of the rescaling variable we fix $m_{c}(m_{c})=1.28$ GeV (close to the world average)
and evaluate the cross sections by exploring five $\lambda$ values (black boxes, with the size of the box increasing with $\lambda$).
Theoretical predictions are better clustered in this case. 
The empty triangle and ellipse indicate central prediction and 90\% C.L. interval based on CT10 NNLO respectively.
The uncertainty of LHC cross sections due to $m_{c}(m_c)$ is comparable to the 
experimental PDF uncertainty and in principle should be included independently from the latter. 

\vspace{\baselineskip}
This work was supported by the U.S. DOE Early Career Research Award
DE-SC0003870 and by Lightner-Sams Foundation.


\end{document}